\documentclass[final,3p,times,twocolumn]{elsarticle}

\usepackage{amssymb}
\usepackage{amsmath}
\usepackage{graphicx}
\usepackage{textcomp}
\usepackage{gensymb}

\usepackage{caption}
\usepackage{subcaption}
\usepackage{hyperref}

\hypersetup{
  colorlinks=true,
  linkcolor=blue, 
  citecolor=red,  
  urlcolor=blue   
}

\journal{European Physical Journal C}

\begin{document}

\begin{frontmatter}

\title{An integrated readout system for parallel-plate avalanche counter and multi-wire drift chamber at HIAF-HIRIBL }

\author[1]{E.~Q.~Liu\corref{cor1}}\ead{liueq@impcas.ac.cn}
\author[1,2]{T.~S.~Huang\corref{cfa1}}\cortext[cfa1]{Co-first author}
\author[1,2]{Z.~X.~Ma}
\author[3,2]{Z.~P.~Sun\corref{cor1}}\ead{sunzp@impcas.ac.cn}\cortext[cor1]{Corresponding authors}
\author[1,2]{L.~Li}
\author[1,2,4,5,6]{H.~J.~Ong\corref{cor1}}\ead{onghjin@impcas.ac.cn}
\author[1,2,4,6]{H.~Wang}
\author[1,5,6]{S.~Terashima}
\author[3,2]{L.~M~Duan}
\author[3,2]{H.~R~Yang}
\author[3,2]{Y.~Qian}
\author[1,2]{F.~S.~Shi}
\author[1,2]{Y.~N.~Song}
\author[7]{B.~H.~Sun}
\author[1,2]{X.~D.~Xu}
\author[3,2]{J.~W.~Yan}
\author[1]{Z.~C.~Zhang}

\address[1]{State Key Laboratory of Heavy Ion Science and Technology, Institute of Modern Physics, Chinese Academy of Sciences, Lanzhou 730000, China}
\address[2]{School of Nuclear Science and Technology, University of Chinese Academy of Sciences, Beijing 100049, China}
\address[3]{Institute of Modern Physics, Chinese Academy of Sciences, Lanzhou 730000, China}
\address[4]{Joint Department for Nuclear Physics, Lanzhou University and Institute of Modern Physics, Chinese Academy of Sciences, Lanzhou 730000, China}
\address[5]{Research Center for Nuclear Physics, Osaka University, Ibaraki, Osaka 567-0047, Japan}
\address[6]{Nishina Center for Accelerator-Based Science, RIKEN, 2-1 Hirosawa, Wako, 351-0198 Saitama, Japan}
\address[7]{School of Physics, Beihang University, Beijing 100191, China}

\begin{abstract}
A newly developed, highly-integrated multi-channel front-end readout system -- FEAM-256 -- is presented for use with position-sensitive gaseous detectors, including parallel-plate avalanche counters (PPACs) and multi-wire drift chambers (MWDCs). 
The system's position resolution was characterized using both an $\alpha$ source and cosmic-ray muons. 
Intrinsic position resolutions of 320 $\micro \rm{m}$ for the PPAC, and 424 $\micro \rm{m}$ for the MWDC were achieved. 
Designed specifically for integration into the data-acquisition infrastructure at the High-Rigidity radioactive Ion Beam Line (HIRIBL) of China's High Intensity heavy-ion Accelerator Facility (HIAF), FEAM-256 enables seamless incorporation of PPAC and MWDC detectors into the HIRIBL experimental setup.
\end{abstract}


\end{frontmatter}


\section{Introduction}
Theoretical calculations predict that approximately 9000 nuclei could be bound~\cite{Xia2018,Toki1995,Mller2016,Goriely2009}. To date, more than 3300 nuclides have been discovered, including 339 naturally occurring nuclides~\cite{Wang2021, Wang2021_1, NUBASE2020, Thoennessen_2018, Thoennessen2016}. Most of these nuclides lie far from the $\beta$-stability valley and are unstable, rendering their properties a subject of intense interest in modern nuclear physics.  This interest has been further driven by the development of radioactive nuclear beam (RNB) facilities~\cite{Ma_PF, Tanihata2016,Gals2003,Blumenfeld2013}. Fragment separators -- such as LISE~\cite{LISE} at Ganil, FRS~\cite{Geissel1992} at GSI, and RIBLL1/RIBLL2~\cite{RIBLL1_Sun2003, RIBLL2_Xu25} at HIRFL, as well as second-generation facilities including Super-FRS~\cite{Geissel2003} of FAIR, BigRIPS~\cite{Kubo2012} at RIKEN, and ARIS~\cite{Hausmann2013} at FRIB -- provide powerful platforms for investigating exotic nuclei and related phenomena using radioactive beams. These facilities enable studies of exotic nuclear structures and responses, short-range correlations, nuclear reaction mechanisms, as well as hypernuclear physics and nuclear astrophysics~\cite{Geesaman2016, Saito2021}.

In-flight fission of $^{238}$U beams and projectile-fragmentation reactions of heavy ion beams have been widely employed at such facilities to produce secondary beams of interest~\cite{Ca60_RIBF,Ge82_MSU,GSI_Fission_1,GSI_Fission_2,RIBF_Fission_1,Mg40_MSU,RIBF_Fission_2}. Particle identification of the secondary beams is achieved using combinations of time-of-flight (TOF), magnetic rigidity ($B\rho$), and energy loss ($\Delta E$), where trajectory reconstruction offers the potential for improved $B\rho$ resolution~\cite{PID_BRS}. Trajectory reconstruction also plays important roles in optimizing transmission, monitoring beam conditions, and reconstruction of reaction vertices. Position detectors such as parallel-plate avalanche counters (PPACs)~\cite{PPAC_RIBF, PPAC_review, DPPAC_JINST} and multi-wire drift chambers (MWDCs)~\cite{CPC_YH,NST_He,Xu2022} have been widely employed owing to their excellent sub-millimeter spatial resolution and radiation hardness under gas-flow operation. 

In China, the High Intensity heavy-ion Accelerator Facility (HIAF)~\cite{Zhou2022} has been constructed, with the first commissioning experiment at the High-Rigidity radioactive Ion Beam Line (HIRIBL, formerly the High-energy Fragment Separator, HFRS~\cite{Sheng2020, Sheng2024}) successfully completed~\cite{Zhang2026}. HIRIBL is expected to open new opportunities for a broad range of nuclear physics experiments in the coming years. 
Given the high counting rates and wide beam distributions expected at the dispersive focal planes of HIRIBL, PPAC detectors with larger sensitive areas and segmented electrodes, and MWDC detectors with smaller wire spacing and larger sensitive areas are required. However, increasing detector size and wire density inevitably leads to a significant increase in the number of readout channels. This poses considerable challenges for experimental preparation, as conventional readout systems comprising separate preamplifiers, shaping amplifiers, discriminators, time-to-digital converters (TDCs) and charge-to-digital converters (QDCs) are not only costly but also require substantial effort to integrate into both the limited onsite experimental space and the HIRIBL data-acquisition (DAQ) system.

Recently, a compact, highly-integrated multi-channel front-end readout electronics board -- featureing a preamplifier, amplifier/shaper, and waveform digitizer -- has been developed and applied to time-projection chambers~\cite{elec_Xu, JINST_QY}. Designated as FEAM-256 and purpose-built for integration into the HIRIBL DAQ system, this board substantially reduces the complexity, cost, and space requirements of detector readout systems, making it an attractive candidate for large-area PPAC and MWDC applications. In this work, we present the spatial-resolution performance of PPACs and MWDCs operated with FEAM-256.

The paper is organized as follows. The details of the electronics board are described in Sec.~\ref{subsec:electronics}. The test setup, including the structures of the PPAC and MWDC detectors, is detailed in Secs.~\ref{subsec:PPAC} and~\ref{subsec:MWDC}, respectively. The analysis methods, including track reconstruction and position-resolution evaluation, are described, and the results are discussed in Sec.~\ref{sec:discussion}. Finally, a summary is given in Sec.~\ref{sec:summary}.

\section{Readout Electronics and Experimental Setup}\label{sec:setup}

\subsection{FEAM-256 Readout Electronics Board} \label{subsec:electronics}
\begin{figure}[!htbp]
    \centering
    \includegraphics[width=\linewidth, trim = 0 0 10 600, clip]{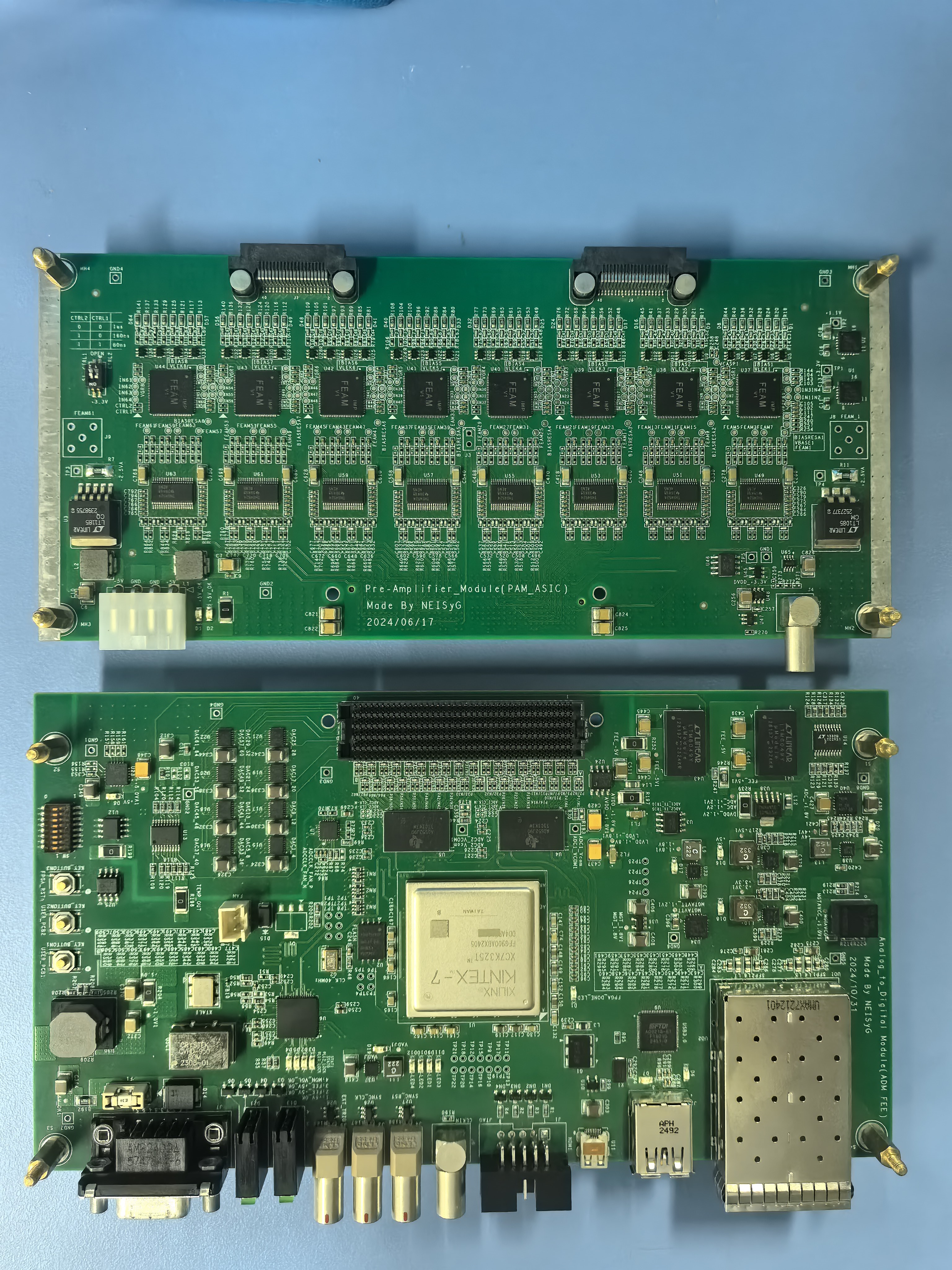}
    \caption{The FEAM-256 readout electronics system, consisting of an ASIC board (top) and a front-end board (bottom) directly connected via a high-density connector.}
    \label{fig:electronics_board}
\end{figure}

The newly developed readout electronics system consists of a dedicated application-specific integrated circuit (ASIC) board and a front-end board, directly interconnected via a high-density connector. Each electronics board supports 64 input channels. On the ASIC board, analog signals from the detectors are amplified and converted from single-ended to differential format. Analog-to-digital conversion and subsequent online processing are then performed on the front-end board, after which the processed data is transmitted for permanent storage. A photograph of the readout electronics is shown in Fig.~\ref{fig:electronics_board}. 

The ASIC board is based on the FEAM (front-end amplifier for MWDC), an 8-channel ASIC. 
Each channel incorporates a charge-sensitive preamplifier, a pole-zero cancellation module, a pulse-shaping circuit, and a driver circuit, providing complete signal processing including amplification, filtering, and shaping of detector output signals. 
The key parameters of the FEAM are summarized in Table~\ref{tab:FEAM_par}. For this application, the amplifier peaking time was set to 80~ns. An overvoltage protection circuit is integrated at the input stage of the ASIC board to prevent accidental damage to the FEAM chip from excessively large-amplitude signals. 
To match the input dynamic range of the analog-to-digital converter (ADC) on the front-end board and fully utilize the ADC sampling precision, the single-ended output signal from the FEAM is amplified by a factor of two using an operational amplifier, and converted to a differential signal before transmission to the front-end board.

\begin{table}[]
    \centering
    \begin{tabular}{cc} \hline
        Number of channels &  8 \\
        Operating voltage & 3.3 V \\
        Input dynamic range & 10 fC -- 1 pC \\
        Peak time & 80~ns, 160~ns, 1~$\mu$s \\
        Gain & 1~mV/fC \\
        Power consumption & 6.6 mW/ch \\
        Integral nonlinearity & $< 1$\% \\
        Noise & 4000 $e^{-}$ @ 100 pF \\ \hline
    \end{tabular}
    \caption{Specifications of the FEAM chip used in this work.}
    \label{tab:FEAM_par}
\end{table}

The front-end board receives differential analog signals from the ASIC board via the high-density connector, and digitizes them using two ADS52J90 ADCs~\cite{ADC}. This ADC supports multiple configurable operating modes; in this work, it is configured for 16-channel input mode with a sampling rate of 80~MSPS and 12-bit resolution. The ADC communicates with the field-programmable gate array (FPGA)~\cite{Xilinx} through the JESD204B protocol~\cite{JESD204}. An LMK04826 phase-locked loop (PLL) chip~\cite{PLL} generates the system clock, the JESD204B reference clock, and the SYSREF synchronization signal required for stable JESD204B operation. The ADC outputs are connected to the GTX gigabit transceiver bank of the FPGA via four current-mode logic (CML) links, each with a maximum bandwidth of 4.8~Gbps. Upon reception by the FPGA, the data undergoes trigger matching, frame formatting and other operations before being uploaded to a computer via a USB~3.0 interface for offline analysis.

The FPGA serves as the core component of the front-end board, undertaking all online data processing tasks. In the FPGA firmware, a data processing time window is triggered immediately upon receipt of an external trigger signal. The window width can be flexibly configured via instruction parameters, ranging from 1 to 4095 ADC sampling cycles. Within the predefined time window, the sampled data is organized into a two-dimensional data frame in chronological and spatial order, while a dedicated frame header containing frame identifier, trigger stamp, timestamp and board number is attached as auxiliary metadata for transmission. To realize precise online matching between trigger signals and particle events, independently adjustable delay chains are implemented for both the trigger signal and the ADC-sampled data in the FPGA firmware. The delay for either the trigger signal or ADC data can be configured via instructions from 0 to 1024 system clock cycles, enabling a maximum matching window of 25.6~$\mu$s under the 80-MHz system clock.

Given that the maximum drift time of the employed detector is approximately 500~ns, the time window width is set to 128 ADC sampling cycles, corresponding to a total duration of 800~ns. With the front-end ADC operating in 16-input-channel mode, a single ASIC board and its matched front-end board handle 32 channels of detector signals. Consequently, the net data payload generated per trigger is 32 (channels) $\times$ 128 (samples) $\times$ 12 (bits) = 6000 bytes, and the framed data is uploaded to the PC for storage via the USB~3.0 interface. The data throughput of the electronics scales linearly with the trigger rate. However, the USB interface does not support long-distance transmission, which limits its applicability in future larger-scale and higher-event-rate scenarios. To address this limitation, a fiber-optic interface with a maximum data bandwidth of 10~Gbps has been reserved on the front-end board to accommodate the high-throughput transmission requirements anticipated for future large-scale system expansion and high-count-rate operations.

In most nuclear physics experiments, position resolutions better than 500~$\mu$m are required.
For the PPAC used in this work, the conversion from the time difference between the two readout sides to position is approximately 0.3~mm/ns. Consequently, a time resolution of 1~ns would yield a position resolution of 300~$\mu$m. 
For the MWDC, the electron drift velocity is typically 0.05~mm/ns. Therefore, an ideal position resolution of 50~$\mu$m could be achieved with a time resolution of 1~ns. 
As discussed in Sec.~\ref{subsec:TimeResBoard}, the intrinsic time resolution of the proposed electronics board satisfies these requirements.

\subsection{Experimental Setup with Parallel Plate Avalanche Counters}\label{subsec:PPAC}
\begin{figure}[t]
    \centering
    \begin{subfigure}{0.55\linewidth}
        \includegraphics[width=\linewidth, trim=50 30 0 10, clip]{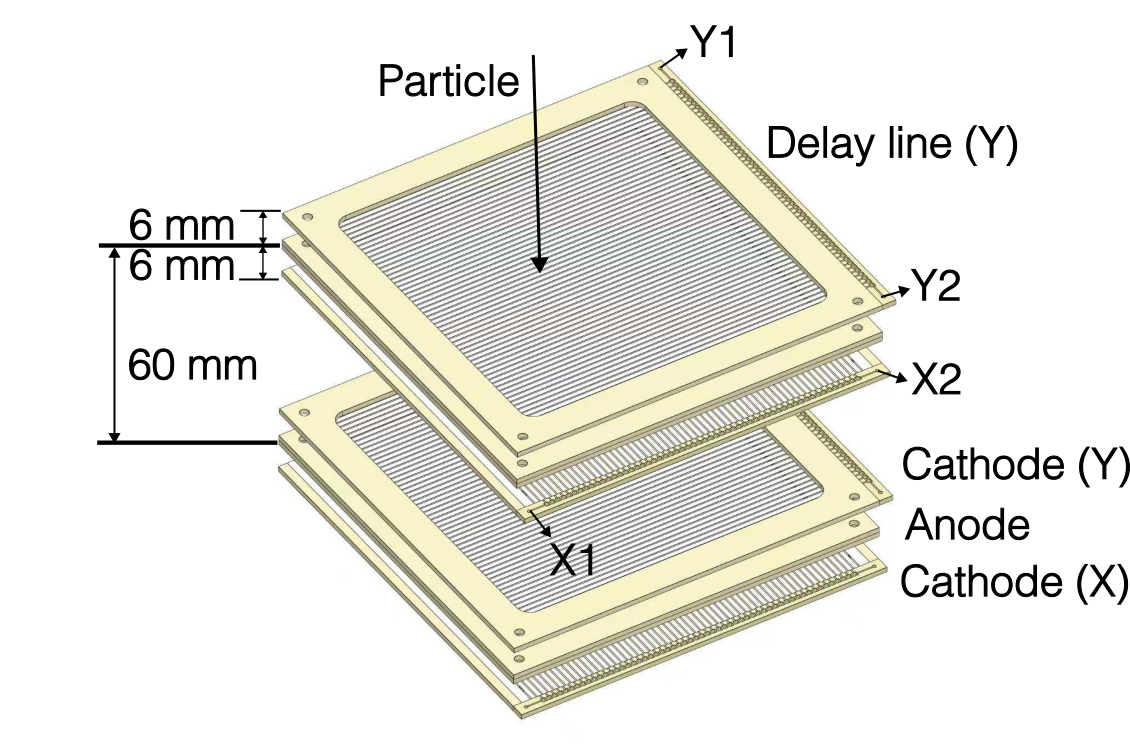}
        \caption{}\label{subfig:PPAC_schematic}
    \end{subfigure}
    \begin{subfigure}{0.43\linewidth}
    \includegraphics[width=\linewidth]{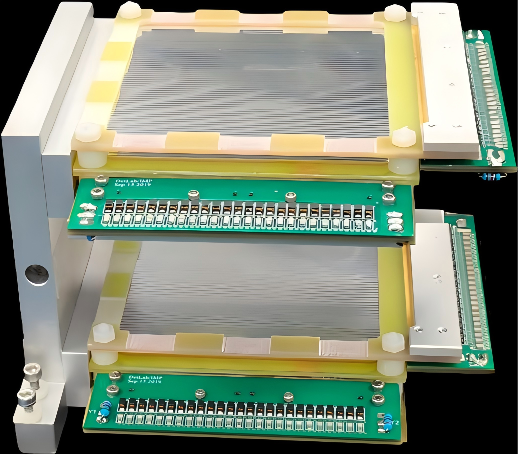}
    \caption{}\label{subfig:PPAC_schematic}
    \end{subfigure}
    \caption{Structure of the DPPAC detector used in this work. (a) Schematic view of the electrode structure. (b) Photograph of the assembled DPPAC. The distance between the two PPACs is 60~mm. The active area is 63~$\times$~63~$\rm{mm^2}$. Adapted from Ref. \cite{DPPAC_JINST}.}
    \label{fig:DPPAC_structure}
\end{figure}

Double parallel-plate avalanche counters (DPPACs) with delay-line readout~\cite{DPPAC_JINST} have been recently developed and applied in secondary-beam experiments at RIBLL1 in HIRFL. 
Each DPPAC consists of two PPACs.
The detector structure is presented in Fig.~\ref{fig:DPPAC_structure}. 
Each PPAC comprises three parallel frames: a central anode flanked by $X$- and $Y$-axis cathode electrodes, ensuring a uniform electric field. The electrodes are fabricated by vacuum evaporation of aluminum onto thin polyester (Mylar) films. The anode is a 2-$\mu$m-thick double-sided aluminized Mylar film sandwiched between two G10 frames. The cathodes feature 1.2-mm-wide aluminum strips with a 1.27-mm pitch and 70-$\mu$m inter-strip gaps, deposited on one side of the film. Each cathode film has 50 strips, providing a sensitive area of 63~mm~$\times$~63~mm. Further details on the detector structure and working principle can be found in Ref.~\cite{DPPAC_JINST}.

In the conventional setup, the timing signals from the anode and cathodes ($T_{X_1}$, $T_{X_2}$, $T_{Y_1}$, $T_{Y_2}$) are processed through a standard readout chain: KB7120 voltage-sensitive preamplifiers (Kaizuworks Co., Ltd.) mounted directly on the DPPAC housing, followed by ORTEC FTA~810 fast-timing amplifiers, constant-fraction discriminators (CFDs), NIM-to-ECL converters, and finally a TDC for digitization. 
In the present work, the signals from the fast-timing amplifiers were routed directly to the FEAM-256 system.

\begin{figure}[htbp]
    \centering
    \includegraphics[width=\linewidth, trim=200 200 50 100, clip]{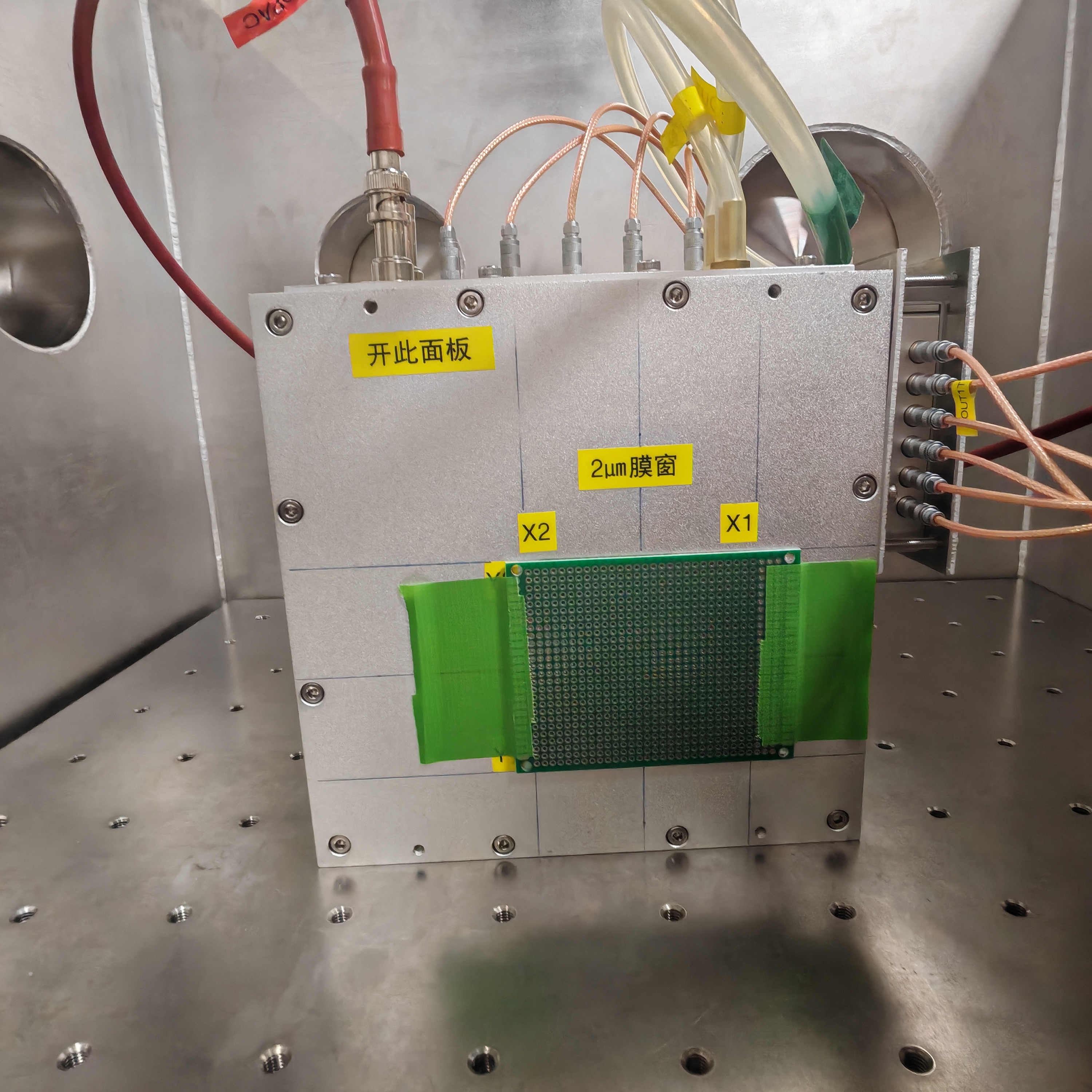}
    \caption{Photograph of the test setup with the DPPAC detector using an $\alpha$ source.  A $^{241}\rm{Am}~\alpha$-source was placed approximately 25 cm in front of the DPPAC detector.}
    \label{fig:PPAC_Setup}
\end{figure}
\begin{figure}[htbp]
    \centering
    \includegraphics[width=\linewidth]{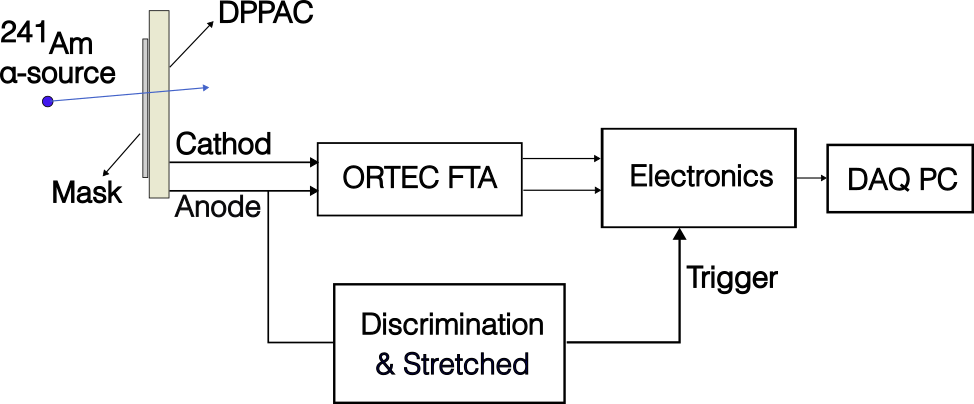}
    \caption{Circuit diagram of the test setup with the DPPAC detector using an $\alpha$ source.}
    \label{fig:PPAC_circuit}
\end{figure}

To evaluate the position resolution of the DPPAC when combined with the new readout electronics, measurements were performed using a $^{241}$Am~$\alpha$ source. 
A schematic view of the test setup and the corresponding circuit diagram are shown in Figs.~\ref{fig:PPAC_Setup} and~\ref{fig:PPAC_circuit}, respectively. The $\alpha$ source was placed approximately 25~cm from the DPPAC detector, and a commercial printed circuit board with a 31~$\times$~26 array of 1-mm-diameter holes was attached to the entrance window of the DPPAC to collimate the incident $\alpha$ particles. With this arrangement, the alignment uncertainty is estimated to be much smaller than the weighted mean widths of the calibration peaks, and therefore does not compromise the measured intrinsic position resolution. 

\subsection{Experimental Setup with Multi-Wire Drift Chambers} \label{subsec:MWDC}
\begin{figure}[!htbp]
    \centering
        \includegraphics[width=\linewidth]{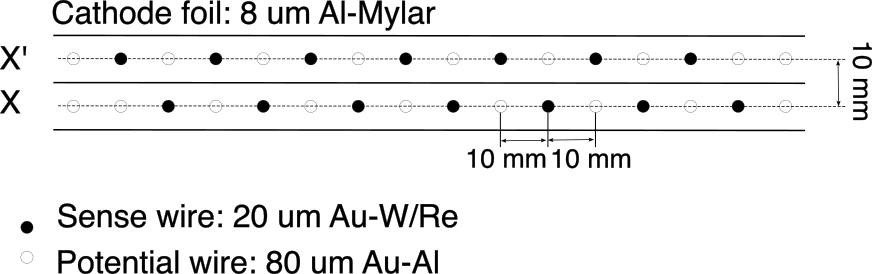}
    \caption{Structure of the multi-wire drift chamber used in this work.}
    \label{fig:MWDC_struct}
\end{figure}

To evaluate the spatial resolution of multi-wire drift chambers (MWDCs) when coupled with the newly developed readout electronics, two MWDCs were installed to measure cosmic-ray trajectories. 
Figure~\ref{fig:MWDC_struct} shows a schematic view of the MWDC used in this work. 
Each MWDC consists of two vertical wire planes and two horizontal wire planes, designated as $XX^{\prime}$ and $YY^{\prime}$, respectively.  Each plane comprises seven sense wires with a pitch of 20 mm, corresponding to a maximum drift length of 10~mm. The primed planes are shifted by half a cell (10 mm) relative to the unprimed planes (see Fig.~\ref{fig:MWDC_struct}), enabling the paired planes to resolve left-right ambiguities and reconstruct hit positions. The sensitive area of each MWDC is 13~cm$\times$13~cm. 
The MWDCs were operated with a continuous flow of a gas mixture of 90\% Ar and 10\% CH$_4$. The sense wires and potential wires were biased at $+1500$~V and $-100$~V, respectively. 

\begin{figure}[htbp]
    \begin{subfigure}{0.50\textwidth}
        \includegraphics[width=\linewidth, trim=0 240 0 240, clip]{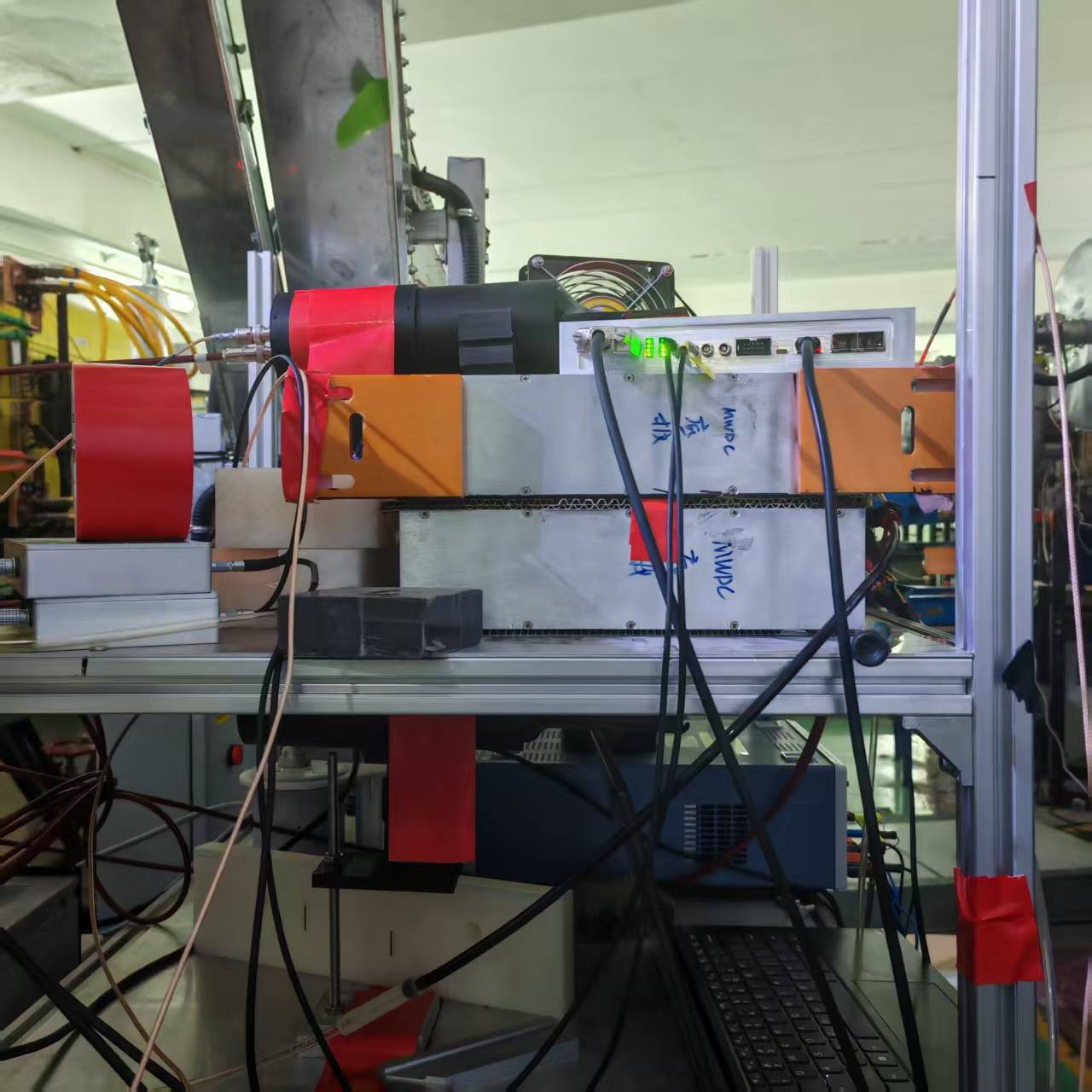}
         \caption{}\label{subfig:photo_MWDC}
    \end{subfigure}
    \begin{subfigure}{0.50\textwidth}
        \includegraphics[width=\linewidth]{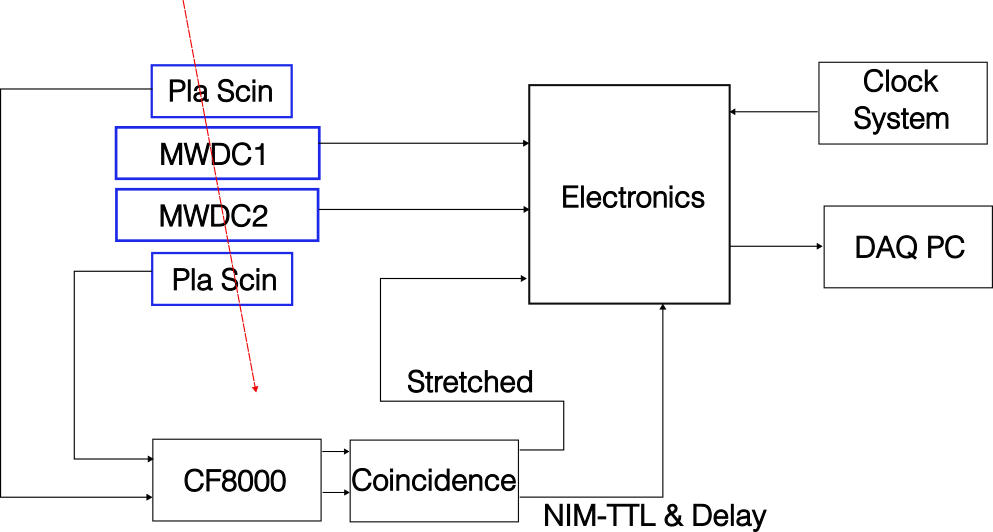}
        \caption{}\label{subfig:circuit_MWDC}
    \end{subfigure}
    \caption{Setup for the cosmic-ray test with MWDCs using the FEAM-256 system. (a) Photograph of the experimental setup. (b) Schematic diagram of the setup and electronics circuit.}
    \label{fig:cosmic_test}
\end{figure}

The two MWDCs were placed horizontally adjacent to each other, with two plastic scintillators coupled to photomultiplier tubes (PMTs) positioned above and below them, respectively. The sizes of the plastic scintillators are $85~\rm{mm} \times 80~\rm{mm} \times5~\rm{mm}$ and $105~\rm{mm} \times 100~\rm{mm} \times4~\rm{mm}$, respectively. The overlap between them is slightly smaller than the sensitive area of the MWDCs.
Signals from the MWDCs were directly fed into the new electronics boards, while a coincidence of the signals from both plastic scintillators served as the trigger for the data acquisition system. This ensured that only events in which cosmic rays penetrated both MWDCs simultaneously were recorded, allowing their trajectories to be reconstructed. A photograph of the experimental setup is shown in Fig.~\ref{subfig:photo_MWDC}. In addition, a stretched trigger signal was also recorded by the same readout electronics as a timing reference; subtracting this reference from the measured timing enabled the elimination of timing jitter, thereby improving the timing resolution. A schematic drawing of the setup and circuit configuration is provided in Fig.~\ref{subfig:circuit_MWDC}.

\section{Results and Discussion} \label{sec:discussion}
\subsection{Intrinsic Time Resolution of the Readout Electronics} \label{subsec:TimeResBoard}
The intrinsic time resolution of the readout electronics was evaluated using an attenuated pulsed signal from a pulse generator. The signal was split and fed into two different channels on the same board. 


\begin{figure}
    \centering
    \includegraphics[width=\linewidth]{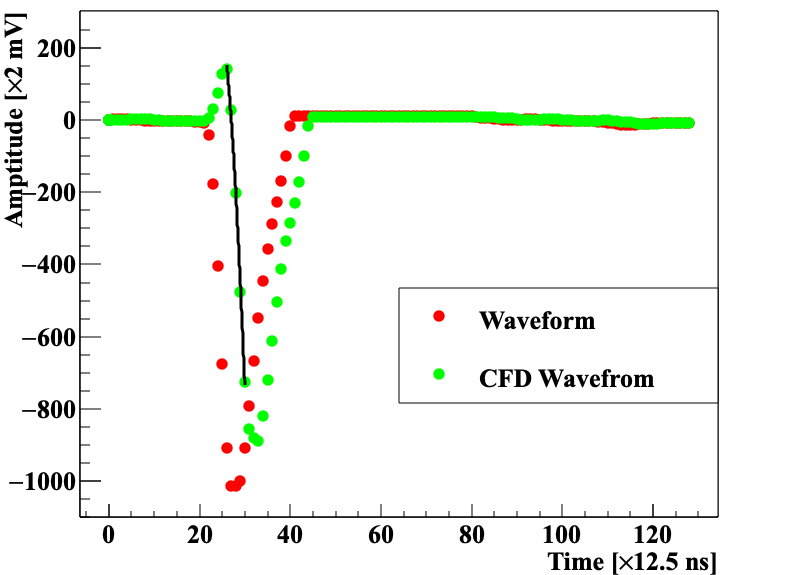}
    \caption{Waveform of cosmic-ray events measured with the MWDC using the new readout electronics.}
    \label{fig:WFD}
\end{figure}
 
The time of each channel was determined using constant-fraction discrimination (CFD) of the waveform.
The CFD waveform is constructed from the sum of the delayed waveform and its inverted, attenuated counterpart:
\begin{equation}
    y(t) = x(t-t_{\rm d}) - \gamma x(t),
\end{equation}
where $y(t)$ is the CFD waveform, $x(t)$ represents the original waveform, $\gamma$ is the attenuation factor, and $t_{\rm d}$ is the applied delay time. The arrival time of the pulse is then determined by fitting the waveform using a second-order polynomial around the zero-crossing point, as illustrated by the black curve in Fig.~\ref{fig:WFD}, taking a waveform from MWDC as an example.

\begin{figure}[t]
    \centering
    \begin{subfigure}{\linewidth}
        \includegraphics[width=\linewidth]{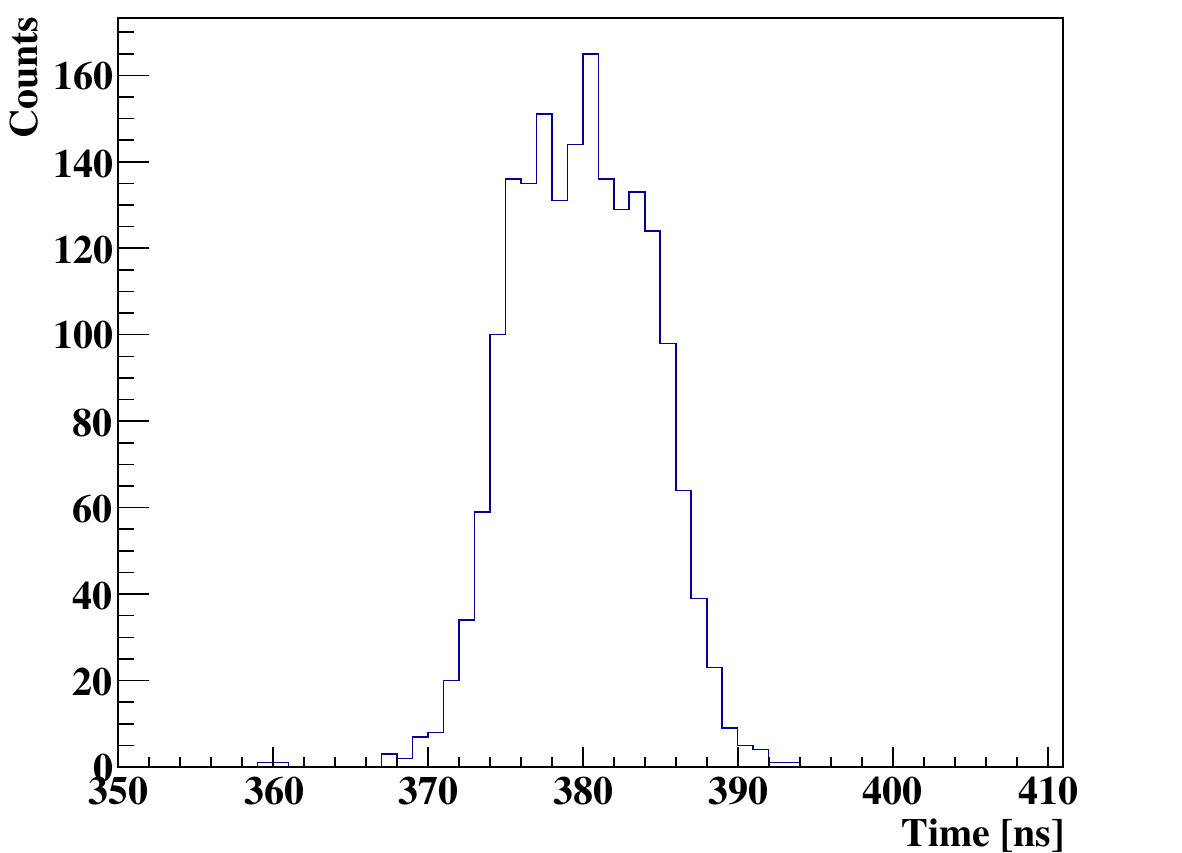}
        \caption{}\label{subfig:TimeDis}
    \end{subfigure}
     \begin{subfigure}{\linewidth}
        \includegraphics[width=\linewidth]{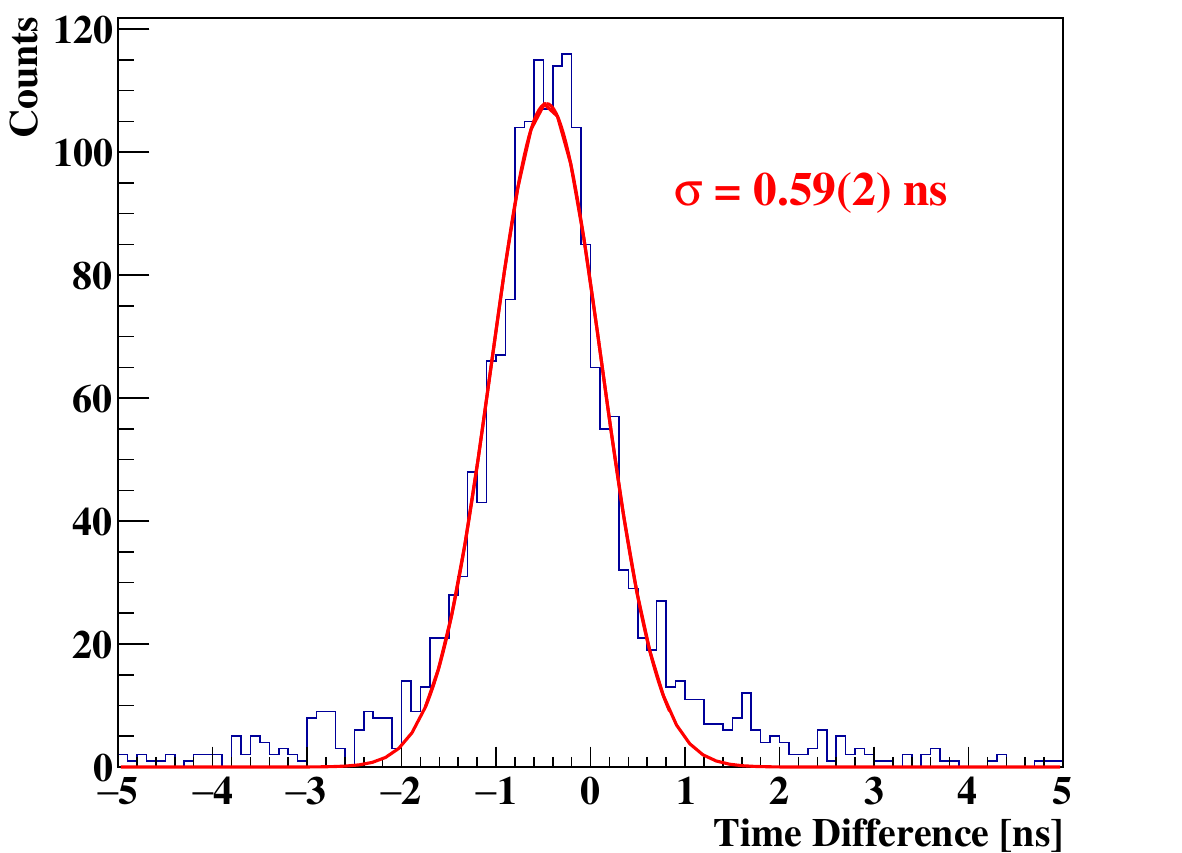}
        \caption{}\label{subfig:TimeDiff}
    \end{subfigure}
    \caption{Distributions of the measured time and the time difference between two channels for signals from a pulse generator. (a) Time distribution for one channel. (b) Time difference between two channels.}
    \label{fig:intrinsicTime}
\end{figure}

Figure~\ref{subfig:TimeDis} shows the measured time distribution for one channel, while Fig.~\ref{subfig:TimeDiff} presents the distribution of the time difference between the two channels. 
Considering the sampling rate of 80 MHz, a 12.5~ns jitter would be included in the determined time for a single channel. Consequently, the time distribution in Fig. \ref{subfig:TimeDis} could be understood by a convolution of an uniform distribution of 12.5~ns and a Gaussian function with a width of $\sigma\sim$1~ns. 
This jitter can be removed by subtracting a reference time measured in the same board. Therefore, it was eliminated in the measured time difference. 
The time-difference distribution was fitted with a Gaussian function within a $\pm2\sigma$ range, as indicated by the red curve in the plot.
The mean value of the time difference indicates offset between different channels.
The intrinsic time resolution, given by the standard deviation of the fit, was determined to be $\sigma = 0.59 \pm 0.02$~ns. This time resolution is sufficient for reading out PPACs and MWDCs. 
It should be noted that unexpected components appear in the tails of the distribution, accounting for approximately 20\% of the total events and slightly degrading the estimated intrinsic time resolution. These components correspond to events with large noise, most likely caused by impedance mismatch at the connection between the input signals and the new electronics.

\subsection{Position Resolution using PPACs}\label{subsec:ana_PPAC}
\begin{figure}[!htbp]
    \centering
    \includegraphics[width=\linewidth]{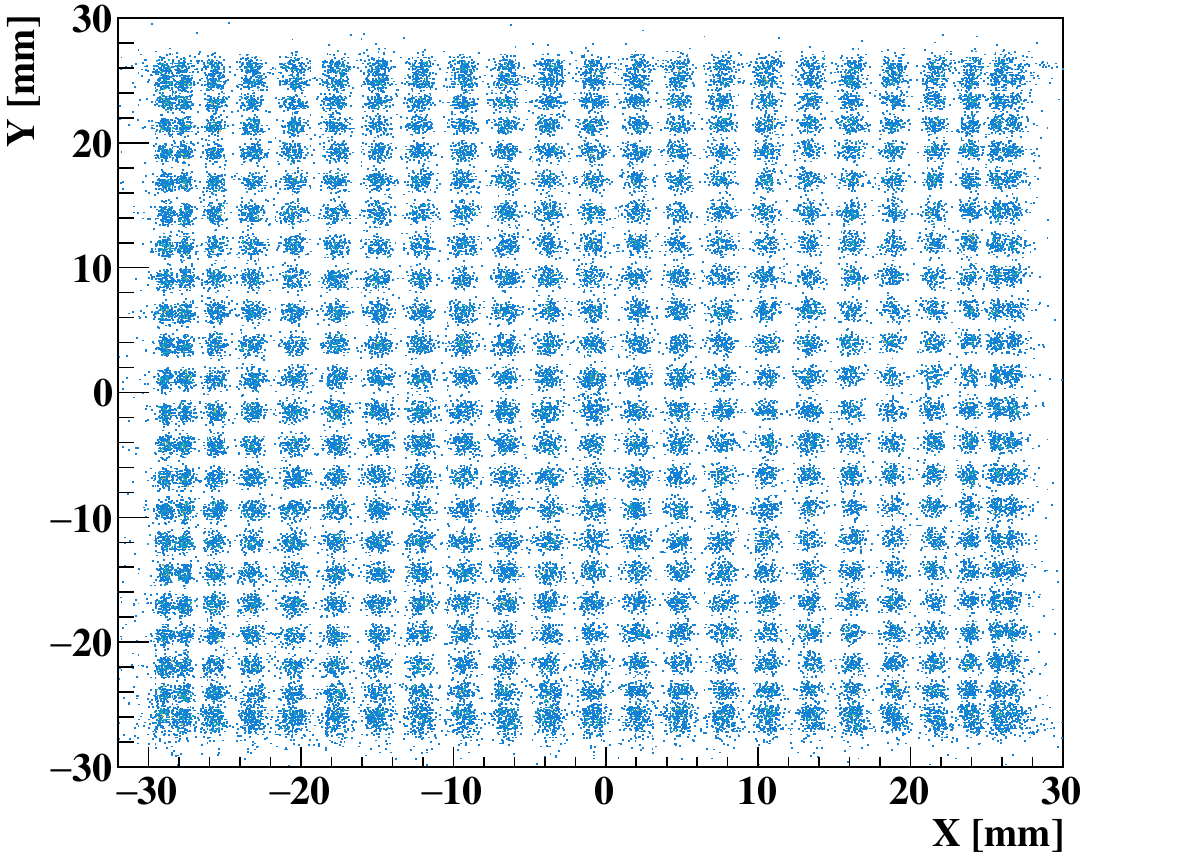}
    \caption{Hit pattern measured by the front PPAC of the DPPAC. A collimating mask was attached to the detector entrance to define the incident particle positions.}
    \label{fig:PPAC_PosDis}
\end{figure}

The measured two-dimensional position distribution of $\alpha$ particles passing through the mask is shown in Fig.~\ref{fig:PPAC_PosDis}. The clearly resolved peaks in the distribution indicate good position resolution. 
It is also observed that the peaks near the edges of the detector differ from those in the center, which is attributed mainly to electron diffusion and, to a lesser extent, to the large-angle incidence of $\alpha$ particles.

\begin{figure}[t]
    \centering
    \begin{subfigure}{\linewidth}
        \includegraphics[width=\linewidth]{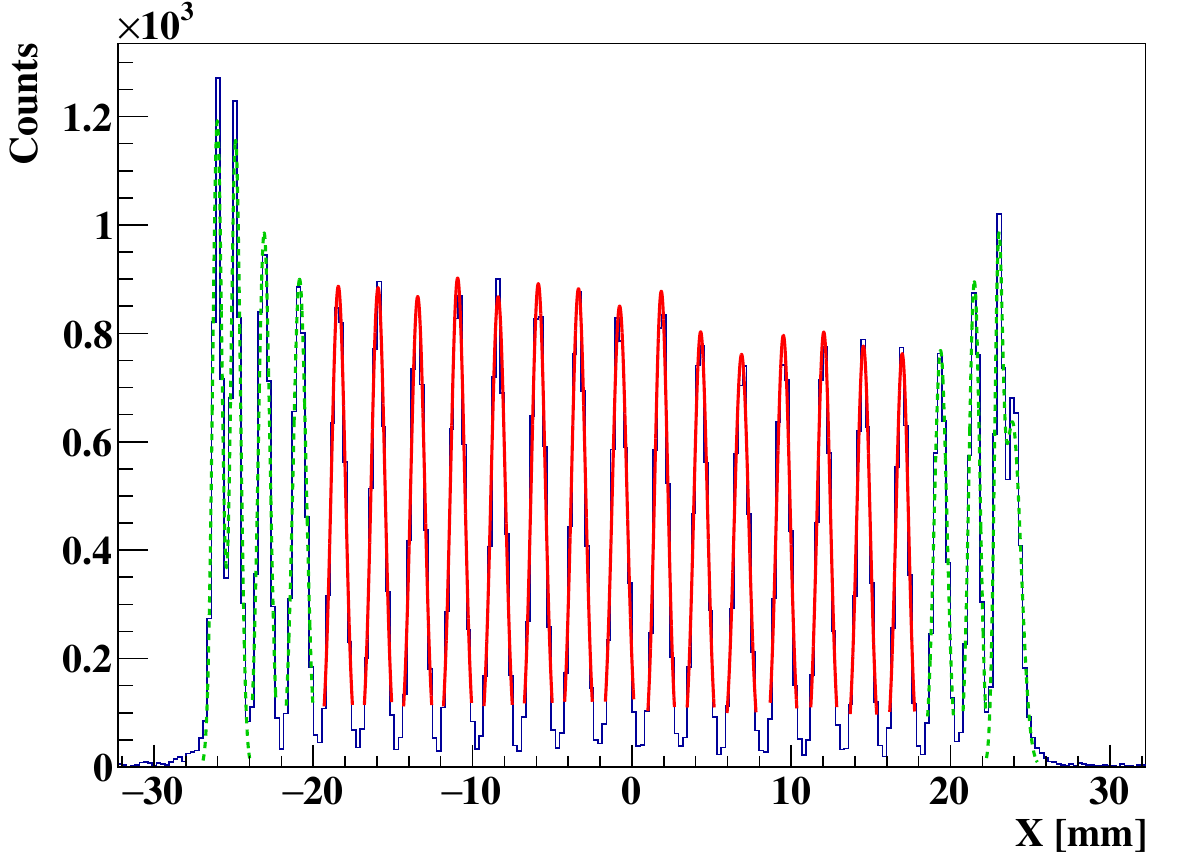}
        \caption{}\label{subfig:PPAC_X}
    \end{subfigure}
    \begin{subfigure}{\linewidth}
        \includegraphics[width=\linewidth]{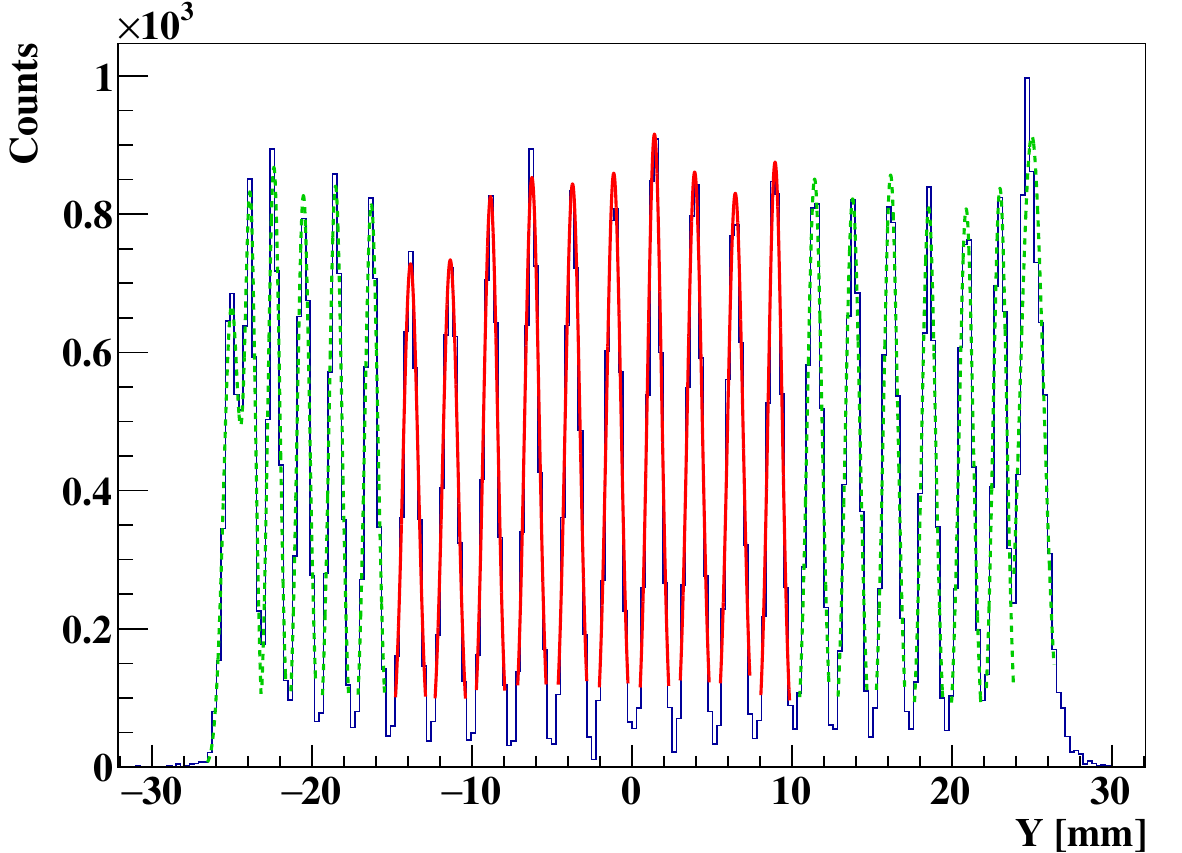}
        \caption{}\label{subfig:PPAC_Y}
    \end{subfigure}   
    \caption{Distribution of collimated hit positions of $\alpha$ particles in the $X$ (a) and $Y$ (b) directions, measured with the mask. The red curves indicate Gaussian fits to each peak; the solid lines denotes those selected to evaluate the resolutions.}
    \label{fig:PPAC_XY}
\end{figure}

To determine the resolution in the $X$ ($Y$) direction, regions containing holes within $-9$~mm~$<Y~(X)<9$~mm were selected and projected onto the $X$ and $Y$ axes, as shown in Figs.~\ref{subfig:PPAC_X} (~\ref{subfig:PPAC_Y}), respectively. 
While the peaks in the central region are distributed uniformly as expected, the pitch between peaks gradually decreases at the edges of the detector because of the electron diffusion mentioned above. Therefore, peaks near the edges of the detector are excluded when determining the resolution in this work. 
The width and associated uncertainty of each peak were extracted by fitting with a Gaussian function. The weighted mean widths for the $X$ and $Y$ directions were determined to be $\sigma_{{\rm meas,}X}= 0.41 \pm 0.02$~mm and $\sigma_{{\rm meas,}Y} = 0.42\pm0.02$~mm, respectively. The uncertainties were evaluated as the weighted mean of the errors of the standard deviations obtained from the Gaussian fits. 

Assuming a uniform distribution of $\alpha$ particles passing through the mask holes, the geometric broadening was determined to be $\sigma_{\rm{geom}}= 0.25$~mm. 
This contribution from the mask holes was corrected using 
\begin{equation}
     \sigma^{2}_{\rm{PPAC}} = \sigma^{2}_{\rm{meas}} - \sigma^{2}_{\rm{geom}},
\end{equation}
The resulting intrinsic position resolutions were obtained as $\sigma_{X} = 0.32\pm0.02$~mm and $\sigma_{Y} = 0.34\pm0.02$~mm (standard deviation), respectively.  

The intrinsic position resolutions achieved with the proposed new readout electronics are comparable to the estimated value presented in Sec.~\ref{subsec:electronics}, and consistent with those obtained with conventional VME-based electronics, as reported in Ref.~\cite{DPPAC_JINST}. Notably, this comparable performance has been achieved with a significantly simplified circuit. The observed position resolution meets the requirements of most nuclear physics experiments, indicating that the proposed readout electronics are suitable for future application with the DPPAC detectors at HIAF-HIRIBL. 

Further developments are in progress. 
Currently, the FEAM-256 system receives signals from the fast-timing amplifiers rather than directly from the KB7120 preamplifiers. We plan to test direct preamplifier readout in the near future. Zero suppression will also be implemented to reduce data size.
Additional experiments are planned to determine the in-beam position resolution and detection efficiency. In addition, the performance and long-term stability of the DPPAC equipped with the new electronics system will be evaluated under high-intensity heavy-ion irradiation, as expected at HIAF-HIRIBL, in the near future.

\subsection{Position Resolution using MWDC}\label{subsec:ana_MWDC}
For cosmic-ray trajectory measurements using the MWDC detectors, a typical waveform generated by a cosmic-ray event is shown in Fig.~\ref{fig:WFD}. The abscissa represents the sampling points of the waveform digitizer, with each sampling point corresponding to 12.5~ns, while the ordinate indicates the signal amplitude in units of 2~mV per channel. Hit information
can be extracted through software analysis of the waveform.
For instance, the hit time was determined using a CFD waveform, as described in Sec. \ref{subsec:TimeResBoard}. 
In the present work, we used $t_{\rm d} = 50$~ns and $\gamma = 0.2$.

\begin{figure}
    \centering
    \includegraphics[width=\linewidth]{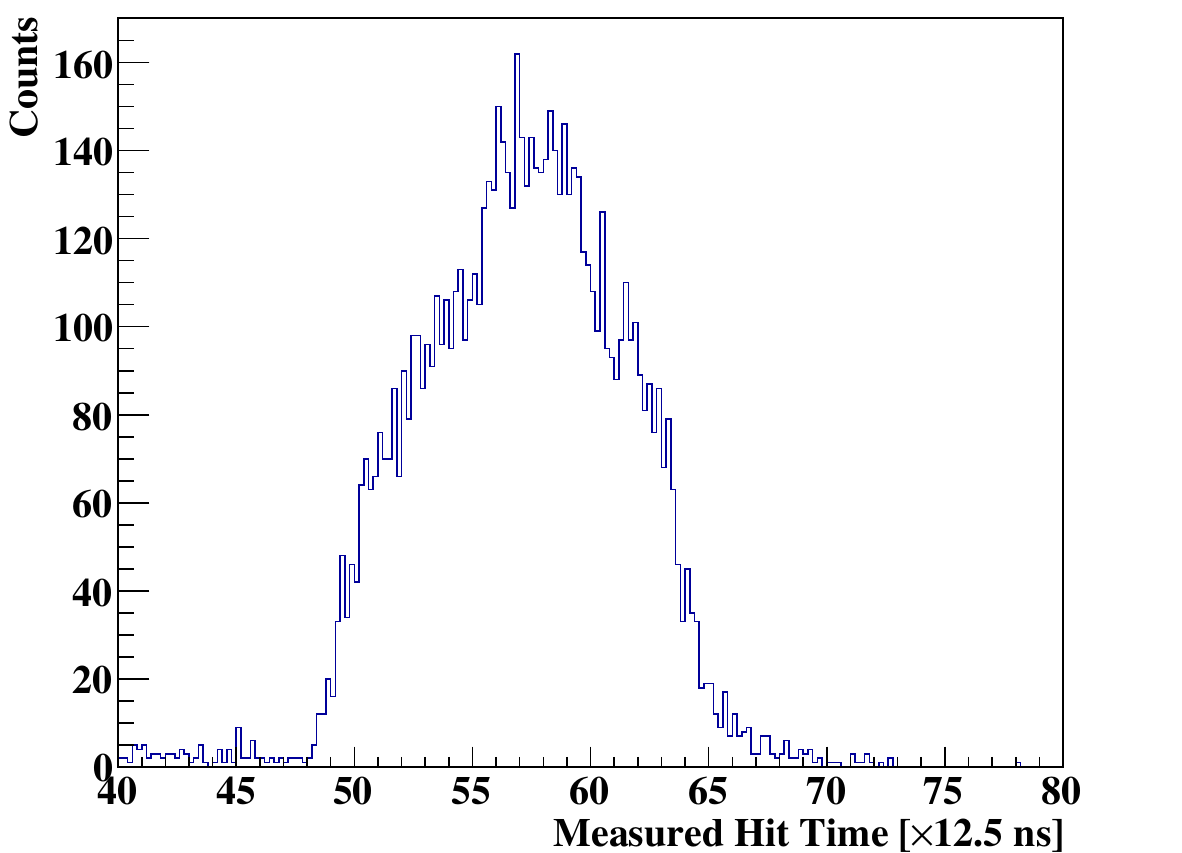}
    \caption{Typical hit time distribution for cosmic rays penetrating the MWDCs.}
    \label{fig:MWDC_TDC}
\end{figure}

A typical hit time distribution after subtracting the reference time for one sense wire is presented in Fig.~\ref{fig:MWDC_TDC}. Additional offsets were applied to align the timing between different channels and to avoid negative values. 
The relationship between the drift length and the measured hit time was established from the hit time distribution using the method described in Ref.~\cite{MWDC_ana}, under the assumption that the drift length is uniformly distributed. 
No obvious drift-time dependence of the residual distribution was observed; therefore, the iterative procedure typically required to improve the resolution was not included in the present analysis. 
The hit position of cosmic rays at each wire plane was then obtained by combining the hit wire position with the evaluated drift length.
Cosmic-ray trajectories were reconstructed using a minimum-$\chi^2$ fitting method, in which the $\chi^2$ function defined by the residuals between the measured hit positions and the predicted track positions was minimized with respect to the track parameters.

\begin{figure}[t]
    \centering
    \begin{subfigure}{\linewidth}
        \includegraphics[width=\linewidth]{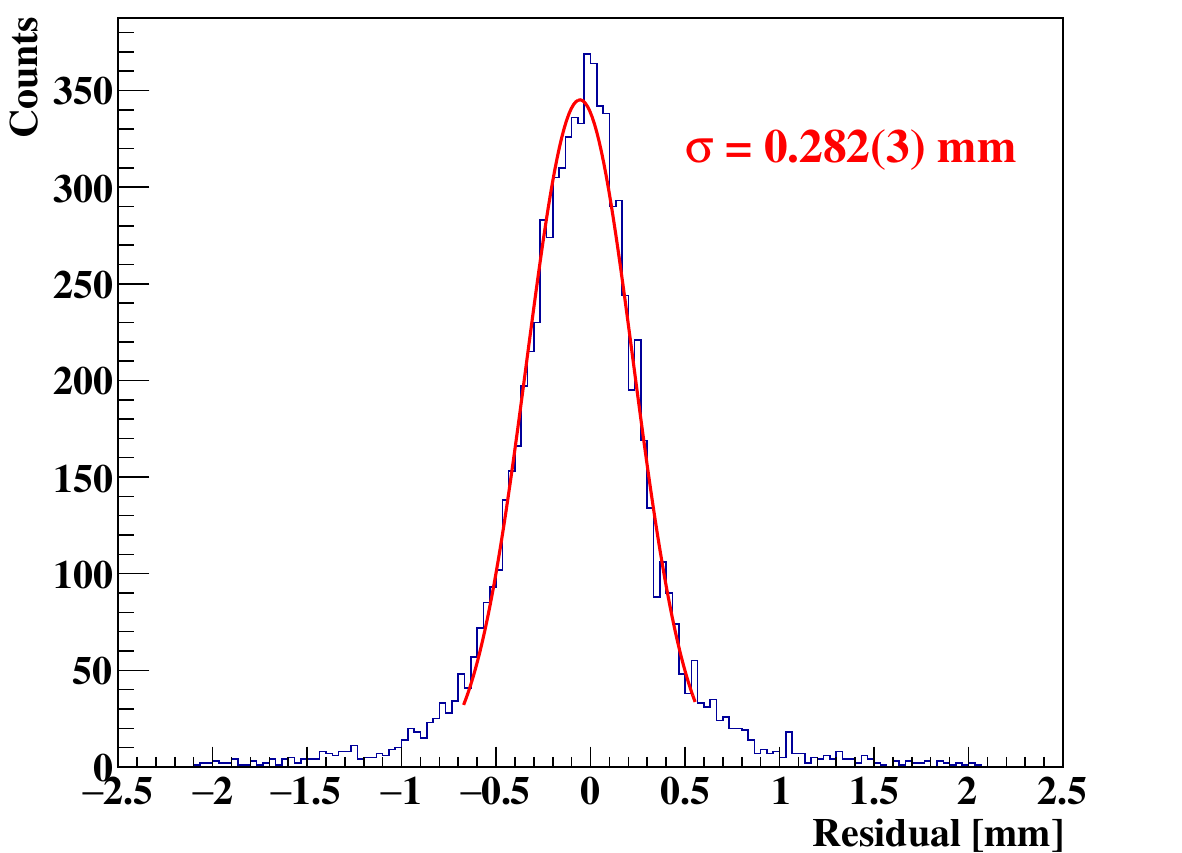}
        \caption{}\label{subfig:MWDC_InRes}
    \end{subfigure}
     \begin{subfigure}{\linewidth}
        \includegraphics[width=\linewidth]{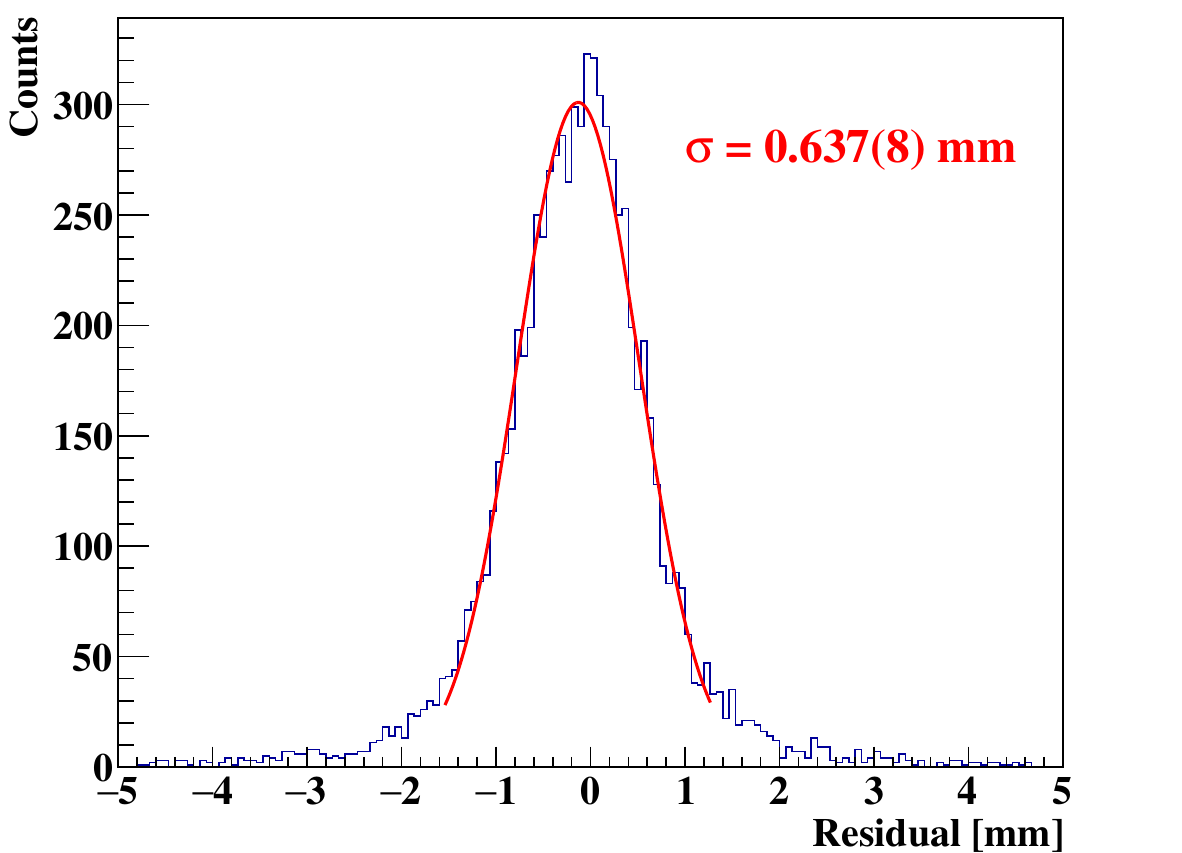}
        \caption{}\label{subfig:MWDC_ExRes}
    \end{subfigure}
    \caption{Residual distributions between the measured hit position and the position predicted by the reconstructed track. The red curves are Gaussian fits within a range of $\pm 2\sigma$.  (a) Residuals using inclusive track reconstruction. (b) Residuals using exclusive track reconstruction.}
    \label{fig:MWDC_res}
\end{figure}

Since the hit positions at each wire plane can be obtained either from the calibrated drift length combined with the nearest sense wire position or from the reconstructed track, the residual between these two determinations serves as a measure of the position resolution. 
The wire plane of interest may be either included in or excluded from the track reconstruction, yielding residuals referred to as inclusive and exclusive residuals, respectively.
For instance, the inclusive and exclusive residual distributions for one wire plane are shown in Figs.~\ref{subfig:MWDC_InRes} and~\ref{subfig:MWDC_ExRes}, respectively. Their widths, extracted from Gaussian fits over a $\pm 2\sigma$ range (red curves), were found to be $\sigma_{\rm in} = 0.282 \pm 0.003$~mm and $\sigma_{\rm ex} = 0.637 \pm 0.008$~mm. The intrinsic position resolution of the plane was then determined as $\sigma = 0.424\pm0.004$~mm using the relation $\sigma = \sqrt{\sigma_{\rm in}\times\sigma_{\rm ex}}$.

The observed position resolution, while worse than that of conventional VME-based systems, meets the requirements of most nuclear physics experiments. A possible cause for this degradation is inadequate grounding or impedance mismatch at the connection between the MWDC and the new electronics. Efforts are ongoing to diagnose and resolve this issue. In addition, the signal-to-noise ratio for the MWDCs was found to be much smaller than that of the PPACs. While the current FEAM chip has a fixed gain of 1~mV/fC, an upgraded version---the FEAT chip---with several selectable gains is expected to resolve this issue. 
Overall, the current test demonstrates the potential of the new readout electronics for future MWDC applications, particularly given its advantages of high channel density, compact size and lower cost. 

\section{Summary} \label{sec:summary}
A highly integrated, multi-channel readout electronics system has been developed, and its spatial-resolution performance has been evaluated in conjunction with parallel-plate avalanche counters (PPACs) and multi-wire drift chambers (MWDCs). Intrinsic position resolutions of approximately 320~$\mu$m and 424~$\mu$m were obtained for the PPAC and MWDC, respectively. These results demonstrate that the new readout electronics, when combined with these detectors, meets the requirements of most nuclear physics experiments. Moreover, the highly integrated design significantly reduces experimental preparation efforts and costs while also alleviating space constraints. 

This work provides a valuable reference for future experiments at the High-Rigidity radioactive Ion Beam Line (HIRIBL) of the High Intensity heavy-ion Accelerator Facility (HIAF) in China. The availability of position-sensitive detectors at each focal plane will greatly facilitate the identification of secondary beams and the optimization of beam transmission through the HIRIBL beamline for a wide range of experiments.

\section*{Acknowledgment}
The authors are grateful to I. Tanihata for supplying the MWDCs used in this work.
This work was supported in part by the National Natural Science Foundation of China (Nos. 12175280, 12175009,  12250610193, 12505146 and 12205348), the Project of Top-notch Leading Talents in Gansu Province, the Research Program of Heavy Ion Science and Technology Key Laboratory, Institute of Modern Physics, Chinese Academy of Sciences (HIST2025KS01), the CAS ``Global Initiative on Common Challenges" under Contract No. 016GJHZ2023063GC, 
the Postdoctoral Research Project of the Department of Human Resources and Social Security of Gansu Province, 
the Longyuan Youth Talent Program of Gansu Province,
Major Science and Technology Projects in Gansu Province under Grant No. 24ZD13GA005 and Technical Innovation Project of Instrument and Equipment Function Development of Chinese Academy of Sciences (Grant No. 2026G102).

\bibliographystyle{elsarticle-num-names} 
\bibliography{reference}

\end{document}